# Chandra X-Ray Observations of PSR J1849-0001, Its Pulsar Wind Nebula, and the TeV Source HESS J1849-000

Seth Gagnon[1], Oleg Kargaltsev[1], Noel Klingler[2,3,4], Jeremy Hare[3,4,5], Hui Yang[1], Alexander Lange[1], and Jordan Eagle[3,6]
[1] Department of Physics, The George Washington University, 725 21st Street, NW, Washington, DC 20052, USA
[2] Center for Space Sciences and Technology, University of Maryland, Baltimore County, Baltimore, MD 21250, USA
[3] Astrophysics Science Division, NASA Goddard Space Flight Center, Greenbelt, MD 20771, USA
[4] Center for Research and Exploration in Space Science and Technology, NASA Goddard Space Flight Center, Greenbelt, MD 20771, USA
[5] The Catholic University of America, 620 Michigan Avenue, N.E. Washington, DC 20064, USA
*Received 2023 November 22; revised 2024 April 10; accepted 2024 April 11; published 2024 June 11*

## Abstract

We obtained a 108 ks Chandra X-ray Observatory (CXO) observation of PSR J1849-0001 and its pulsar wind nebula (PWN) coincident with the TeV source HESS J1849-000. By analyzing the new and old (archival) CXO data, we resolved the pulsar from the PWN, explored the PWN morphology on arcsecond and arcminute scales, and measured the spectra of different regions of the PWN. Both the pulsar and the compact inner PWN spectra are hard with power-law photon indices of $1.20 \pm 0.07$ and $1.49 \pm 0.20$, respectively. The jet-dominated PWN has a relatively low luminosity, lack of $\gamma$-ray pulsations, relatively hard and nonthermal spectrum of the pulsar, and sine-like pulse profile, which indicates a relatively small angle between the pulsar's spin and magnetic dipole axis. In this respect, it shares similar properties with a few other so-called MeV pulsars. Although the joint X-ray and TeV spectral energy distribution can be roughly described by a single-zone model, the obtained magnetic field value is unrealistically low. A more realistic scenario is the presence of a relic PWN, no longer emitting synchrotron X-rays but still radiating in TeV via inverse-Compton upscattering. We also serendipitously detected surprisingly bright X-ray emission from a very wide binary whose components should not be interacting.

*Unified Astronomy Thesaurus concepts:* Pulsars (1306); Rotation powered pulsars (1408); Pulsar wind nebulae (2215); Gamma-ray sources (633); X-ray astronomy (1810)

## 1. Introduction

Pulsars are among nature's most powerful particle accelerators, capable of producing particles with energies up to a few PeV. As a pulsar spins down, most of its rotational energy is converted into a magnetized ultrarelativistic-particle wind, whose synchrotron emission can be seen from radio to $\gamma$-rays as a pulsar wind nebula (PWN). To date, the largest number of PWNe have been discovered in X-rays with Chandra X-ray Observatory (CXO hereafter; see Kargaltsev & Pavlov 2008; Kargaltsev et al. 2013, 2017; Reynolds et al. 2017 for reviews), thanks to its unrivaled angular resolution and the very low ACIS background. For many PWNe, the sharp CXO images show that the pulsar winds are highly anisotropic and dominated by equatorial and polar outflow components (tori and jets, respectively), which reflect the rotational symmetry of the pulsar (Kargaltsev & Pavlov 2008). These structures not only tell us about the properties of the wind (e.g., the PWN surface brightness and spectrum can be used to constrain the magnetic field strength and electron energies) but also carry an imprint of the intrinsic properties of the pulsar (e.g., the pulsar velocity, and the spin and magnetic axes orientation; Bühler & Giomi 2016; Kargaltsev et al. 2017).

Developments in modeling pulsar magnetospheres allow one to constrain the viewing angle ($\zeta$) and magnetic inclination angle ($\alpha$) from the radio and $\gamma$-ray pulse profiles (e.g., Watters et al. 2009; Pierbattista et al. 2015). However, some pulsars are quiet in radio and GeV $\gamma$-rays, thus requiring a different approach. Ng & Romani (2004, 2008) have demonstrated that even faint compact PWN structures (tori and jets) can be used to infer the 3D orientation of the pulsar spin axis once Doppler boosting is taken into account. Obtaining independent constraints on $\zeta$ and $\alpha$ from PWN morphologies provides an important check for pulsar magnetospheric models (Bai & Spitkovsky 2010; Romani & Watters 2010; Cerutti et al. 2016; Pierbattista et al. 2016).

Fitting PWN jets/tori (Ng & Romani 2004, 2008) can also provide an accurate measurement of the termination shock scale for the equatorial wind and the relative luminosities of the tori and jet components (possibly related to the angle $\alpha$ between the magnetic and spin axis; Bühler & Giomi 2016). For example, the relative strengths of the jets and tori can vary: some PWNe appear to be dominated by equatorial outflows (e.g., the Crab, Vela, 3C58, and G54.1+0.3), while others are dominated by jets (e.g., B1509-58, Kes 75, and G11.2–0.3). Also, some PWNe exhibit a single torus (e.g., the Crab; Weisskopf et al. 2000), while others show double tori (e.g., Vela and the "Dragonfly Nebula"; Pavlov et al. 2003 and Van Etten et al. 2008, respectively). Finally, measuring the angle between the pulsar velocity vector and spin axis sheds light on the supernova (SN) explosion processes that result in kicks to the neutron stars (Ng & Romani 2007). For these reasons it is important to continue sensitive, high-resolution X-ray observations of PWNe.

The INTEGRAL source IGR J18490–0000 was discovered during a survey of the Sagittarius and Scutum arms (Stephen et al. 2006). An observation with XMM-Newton revealed the

---

[6] NASA Postdoctoral Program Fellow.

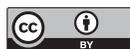

**Table 1**
Observed and Derived Pulsar Parameters

| Parameter | Value |
| --- | --- |
| R.A. (J2000.0) | 18:49:01.632 (40) |
| decl. (J2000.0) | –00:01:17.45 (60) |
| Epoch of Position (MJD) | 55,885 |
| Galactic Longitude (deg) | 32.64 |
| Galactic Latitude (deg) | 0.53 |
| Spin Period, $P$ (ms) | 38.52 (15) |
| Period Derivative, $\dot{P}$ ($10^{-15}$) | 14.16 (11) |
| Surface Magnetic Field, $B_S$ ($10^{11}$ G) | 7.47 |
| Spin-down Power, $\dot{E}$ ($10^{36}$ erg s$^{-1}$) | 9.8 |
| Spin-down Age, $\tau_{sd} = P/(2\,\dot{P})$ (kyr) | 43.1 |

**Note.** Parameters are from the ATNF Pulsar Catalog (Manchester et al. 2005). Values in parentheses are the 1σ uncertainties.

pulsar/PWN nature of this source (Terrier et al. 2008). Gotthelf et al. (2010) used RXTE observations to discover an energetic pulsar, PSR J1849–0001, with period $P = 38.5$ ms, a spin-down energy-loss rate $\dot{E} = 9.8 \times 10^{36}$ erg s$^{-1}$, and characteristic age $\tau_c = 42.9$ kyr. This pulsar has the 16th highest $\dot{E}$ out of the ∼2400 pulsars known in our Galaxy (Manchester et al. 2005), and hence, it must be relatively young, even if its characteristic age exceeds its true age (see, e.g., Suzuki et al. 2021). The RXTE pulse profile shows a single broad peak (FWHM ≃ 0.35 pulse phase) with a small dip at the maximum. See Table 1 for all pulsar parameters. Gotthelf et al. (2010) fitted the XMM-Newton EPIC spectrum of the bright central source (including an unresolved compact PWN) and found that it can be described by an absorbed power law (PL) with a hydrogen column density $n_{\rm H} = (4.3 \pm 0.6) \times 10^{22}$ cm$^{-2}$, photon index $\Gamma = 1.1 \pm 0.2$, and observed (absorbed) flux $f_{2-10{\rm keV}} = (3.8 \pm 0.3) \times 10^{-12}$ erg cm$^{-2}$ s$^{-1}$. The large $n_{\rm H}$ suggests a distance of about 7 kpc, placing the pulsar within the Scutum tangent region. In addition, a short 23 ks Chandra observation was obtained in 2012 (ObsID 13291) to explore the compact structure of the PWN. Kuiper & Hermsen (2015) and Kim et al. (2024) used this observation to analyze the PSR J1849–0001 and its PWN spectra. For the PWN,[7] Kuiper & Hermsen (2015) fitted an absorbed PL model, fixing $n_{\rm H}$ to $4.5 \times 10^{22}$ cm$^{-2}$, and found a photon index of $\Gamma = 1.18 \pm 0.05$ and an unabsorbed flux $f_X = (0.71 \pm 0.05) \times 10^{-12}$ erg cm$^{-2}$ s$^{-1}$ in 2–10 keV. For the pulsar spectrum fitted with absorbed PL, Kuiper & Hermsen (2015) obtained the best-fit $\Gamma = 1.08 \pm 0.02$ for the same fixed $n_{\rm H}$ and the unabsorbed flux of ≈$4.1 \times 10^{-12}$ rg cm$^{-2}$ s$^{-1}$ in 2–10 keV. Kim et al. (2024) fit the PWN with a PL + logpar model with a fixed $n_{\rm H} = 6.4 \times 10^{22}$ cm$^{-2}$ and found $\Gamma = 1.96 \pm 0.33$ and an unabsorbed flux $f_X = (1.44 \pm 0.18) \times 10^{-12}$ erg cm$^{-2}$ s$^{-1}$ in 2–10 keV. In this study, we analyzed the data from the first short CXO observation together with the data from our new CXO observations.

Many young pulsars with X-ray PWNe are also accompanied by the TeV sources (see, e.g., Kargaltsev et al. 2013; H.E.S.S. Collaboration et al. 2018a). The H.E.S.S. observations revealed a slightly extended TeV source, HESS J1849–000, positionally coincident (within the HESS source uncertainty; H. E. S. S. Collaboration et al. 2018b) with IGR J18490–0000 and PSR J1849–0001 (which are the same source). The TeV emission was also detected at higher energies by HAWC (Abeysekara et al. 2017), but the 3HWC J1849+001 position is offset by ∼13′ from the pulsar position. Recently, unpulsed γ-rays were detected up to 320 TeV by the Tibet air shower array (TASA) making the source a Pevatron candidate (Amenomori et al. 2023). The pulsar and PWN are not, however, detected in GeV γ-rays with Fermi-Large Area Telescope (Fermi-LAT). There is no known radio supernova remnant (SNR) coincident with HESS J1849–000 (Anderson et al. 2017; Green 2019).

The paper is organized as follows. In Section 2 we describe the data and data reduction. In Section 3 we present the results of the data analysis. In Section 4 we discuss the results, and in Section 5 we conclude with a summary.

## 2. Data

We have obtained three Chandra observations of PSR J1849–0001 (ObsIDs 23596, 24494, and 24495). These observations occurred on 2022 September 15, 2021 September 26, and 2022 September 11 with exposures of 49.41, 29.67, and 29.67 ks, respectively. The data were collected using Advanced CCD Imaging Spectrometer (ACIS; Garmire et al. 2003) I-array operated in the "Very Faint" timed exposure mode. The total scientific exposure was 108.75 ks. We also analyzed the archival 22.7 ks observation (ObsID 13291; PI: Kaastra) carried out with the ACIS S-array on 2012 November 16. The target imaged on the back-illuminated S3 chip operated in the "Faint" timed exposure mode.

The data from these observations were reprocessed using the Chandra Interactive Analysis of Observations (CIAO) software package (Fruscione et al. 2006) version 4.15 with the Chandra Calibration Database version 4.10.4. The most recent calibrations were applied to the data using the `chandra_repro` tool.

We performed a correction of the World Coordinate System (WCS) information in all four observations by coaligning the reference X-ray sources from these observations to their Gaia Data Release 3 (DR3) counterparts (Gaia Collaboration et al. 2023). The details of this procedure are given in the Appendix. The event lists from these four observations were then merged using `merge_obs` and restricted to the 0.5–7.0 keV energy range.

The spectral fits were performed in Sherpa (v.4.15.0; Freeman et al. 2001) using the XSPEC absorbed PL model, `xstbabs.gal* xspowerlaw.pl;` (the absorption cross sections and abundances are described in Wilms et al. 2000). The best-fit model parameter uncertainties are given at the 68% (1σ) confidence level.

## 3. Results

### 3.1. Images

Figure 1 shows the X-ray view of the HESS J1849–000 field as seen by ACIS-S and ACIS-I observations. The PWN is better discernible in the combined ACIS-I image because of the longer exposure and lower ACIS-I background. The merged ACIS-I image reveals a compact PWN consisting of an elongated structure extending to the southwest of the pulsar, likely a pulsar jet exhibiting helicity. In addition to the bright compact PWN (the jet; see the inset in Figure 1), fainter and more extended PWN emission is seen northeast of the pulsar.

---

[7] To analyze the PWN spectrum, both studies chose regions similar to our Region C (see Figure 2).

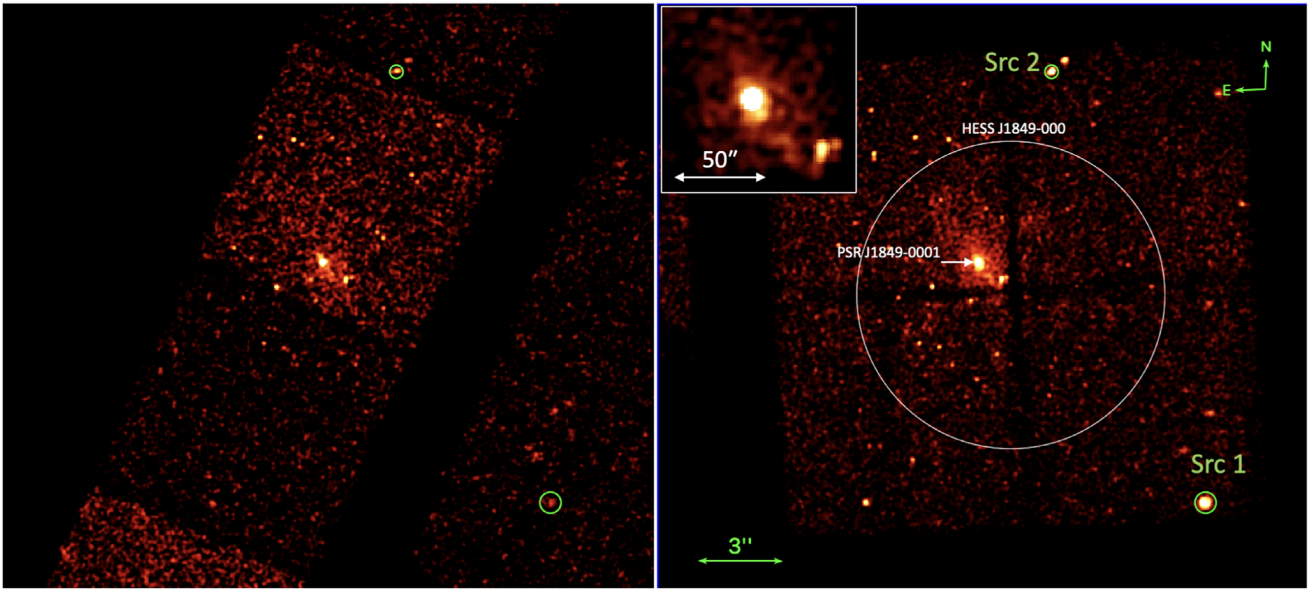

**Figure 1.** The ACIS-S (left; ObsID 13291) and ACIS-I (right; ObsIDs 23596, 24494, and 24495 combined) images of the HESS J1849−000 field (the size of the TeV source is shown by the white circle) in 0.5–7 keV. The inset shows a zoomed-in view of the brighter part of PWN powered by PSR J1849−0001. Two green circles labeled Src 1 and Src 2 show the two brightest (in the combined ACIS-I image) sources.

A number of point sources are seen in the images with the two most prominent (brightest) ones (Source 1 = CXOU J184829.9−000943 and Source 2 = CXOU J184851.2+000523) marked by the green circles in the ACIS-I image. We note that Source 1 is strongly variable.

The top panel in Figure 2 shows the image produced by applying exposure-map corrections and subtracting point sources. CIAO's `wavdetect` (Freeman et al. 2002) was used to detect the point sources, and their respective background regions were created by `roi`. The point sources were then replaced with background regions using `dmfilth`, creating a diffuse emission image. The exposure correction was performed by dividing the diffuse emission image by the exposure map with `dmimgcalc`. The individual images and exposure map were binned by a factor of 2. The resulting images from each observation are then combined (using `dmimgcalc`), and the result was smoothed with `aconvolve`. The putative helical jet extending SW of the pulsar is more apparent in this exposure-corrected diffuse emission image.

### 3.2. Pulsar Motion

Once the astrometry in the individual observations was corrected, as described in the Appendix, we measured the proper motion from the images. We used a standard centroiding approach (described in, e.g., Da Costa 1992) to measure pulsar positions for $r = 0.\!\!^{\prime\prime}37$, $0.\!\!^{\prime\prime}49$, and $0.\!\!^{\prime\prime}61$ circular apertures[8] in ObsID 13291 and 24494, which are separated by 3590 days. These two positions differ by $-0.\!\!^{\prime\prime}16$ in R.A. and $-0.\!\!^{\prime\prime}03$ in decl. Since this difference is noticeably smaller than the weighted root-mean-square residuals (WRMSRs) after the astrometric alignment (see Table 3 in the Appendix), we conclude that the pulsar's proper motion could not be detected with the existing data. We note that the WRMSR of $0.\!\!^{\prime\prime}18$ (see Appendix) corresponds to ≈600 km s$^{-1}$ at the fiducial distance of 7 kpc. This can be taken as an upper limit on pulsar's velocity.

### 3.3. PWN and Pulsar Spectra

The bottom panel in Figure 2 shows regions used for spectral extraction on top of the combined (from all four observations) ACIS image. The pulsar's spectrum was extracted from $r = 3.\!\!^{\prime\prime}0$ circular aperture centered at the brightest pixel (region A in Figure 2, lower panel), which corresponds to the ≈90% encircled point-spread function (PSF) fraction at 1.5 keV. This radius was chosen by studying the pulsar contribution to the inner PWN region. The net counts in the inner PWN region in the X-ray image were compared to the net counts in that region from a point source simulated using MARX at varying radii (Davis et al. 2012). A "buffer" region was introduced between the pulsar and inner PWN regions to reduce the contamination within the inner PWN region from the pulsar. Comparing the CXO observations and the MARX simulation, ≈7% of the net counts in the inner PWN region may be attributed to the pulsar.

To fit the pulsar spectrum (shown in Figure 3), we first grouped counts by requiring ⩾30 counts per bin and restricted the energy range to 0.5–7.0 keV. The PL model, modified by interstellar absorption, provides a good fit with a reduced $\chi^2$ $\chi_\nu^2 = 1.03$ for $\nu = 311$ degrees of freedom (d.o.f.). The best-fit PL $\Gamma = 1.20 \pm 0.07$, while $n_H = (4.40 \pm 0.18) \times 10^{22}$ cm$^{-2}$.[9] The corresponding absorbed and unabsorbed fluxes are $(2.0 \pm 0.2) \times 10^{-12}$ erg cm$^{-2}$ s$^{-1}$ and $3.9^{+1.3}_{-1.5} \times 10^{-12}$ erg cm$^{-2}$ s$^{-1}$, respectively. The unabsorbed luminosity at the assumed $d = 7$ kpc is $(2.3 \pm 0.1) \times 10^{34}$ erg s$^{-1}$. The $n_H$ inferred from the PL fit to the pulsar's spectrum is used in the PWN fits below.

The effects of pileup were studied using XSPEC's `pileup` model (Davis 2001). We performed multiple fits while varying the $\alpha$ parameter of this model to understand its impact on our absorbed PL model fit. We found only slight deviations compared to fitting without pileup. In particular, the photon index slightly softened to a value $\Gamma \approx 1.25 \pm 0.08$, $n_H$ increased to $n_H \approx (4.56 \pm 0.22) \times 10^{22}$ cm$^{-2}$, and the normalization became $\mathcal{N}_{-5} \approx 52.18 \pm 6.58$ in units of $10^{-5}$ photons s$^{-1}$ cm$^{-2}$ keV$^{-1}$ at 1 keV. These changes are within the uncertainties of the values without pileup; therefore, our conclusion is that the effects of pileup are quite modest for this source, which is in agreement with the prior work of Kuiper & Hermsen (2015).

---

[8] We use small apertures to avoid the known asymmetry in the PSF anomaly located ≈$0.\!\!^{\prime\prime}8$ from the centroid (see https://cxc.harvard.edu/proposer/POG/html/chap4.html).

[9] Although this value is close to the one reported in Gotthelf et al. (2010), the latter value was measured from a different region, and it is unclear what abundances were used in that work. Therefore, the apparent consistency should be taken with caution.

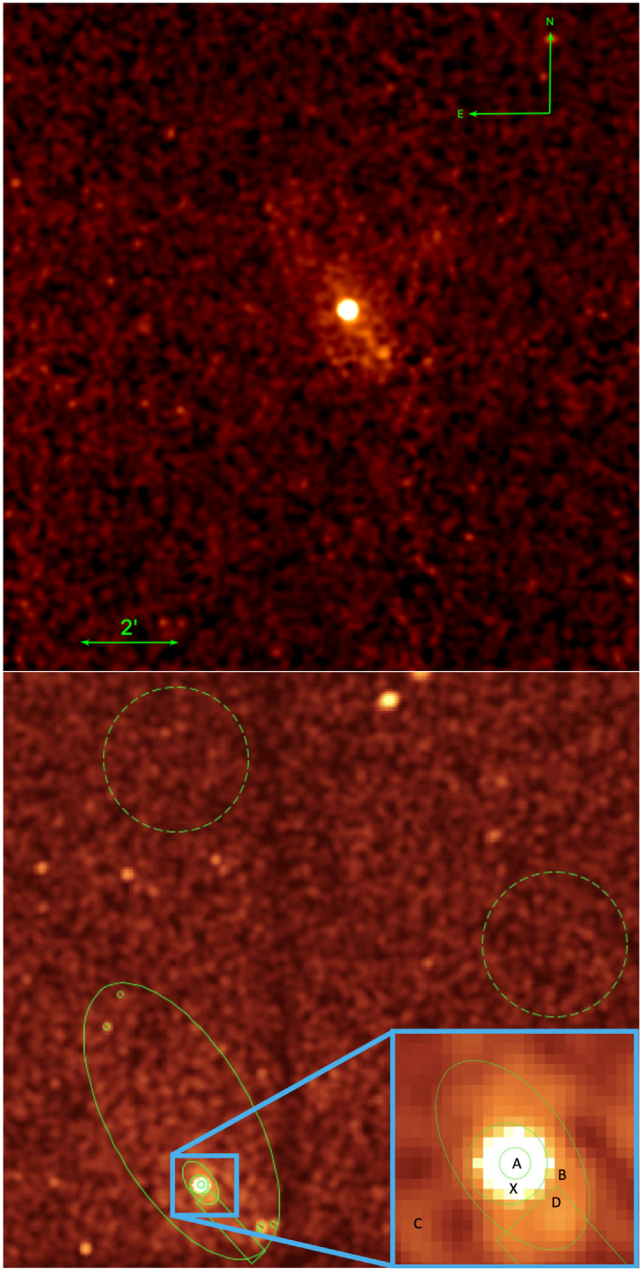

**Figure 2.** The top panel shows the point-source-subtracted and exposure-map corrected image of the J1849 PWN. The image is binned by a factor of 4 and smoothed with a 2 pixel Gaussian kernel. The bottom panel shows regions used for spectral extractions, with the background regions indicated by dashed lines: A–pulsar ($r = 3.''0$), B–inner PWN, C–extended PWN, D–jet candidate, and X–buffer region excluded from the analysis.

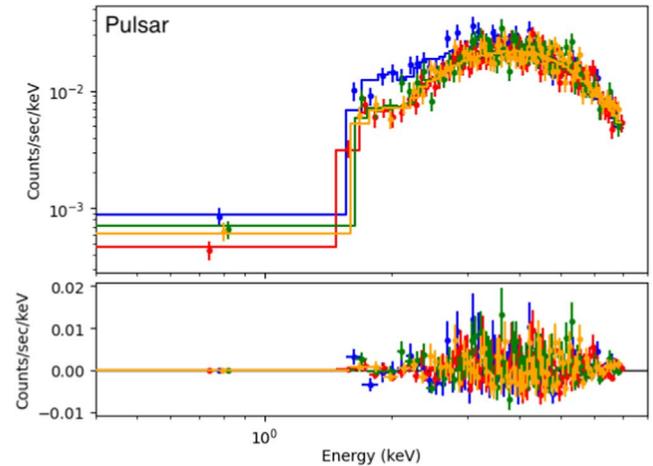

**Figure 3.** Spectrum of PSR J1849-0001. Different colors represent each observation (blue–13291, red–23596, green–24494, and orange–24495.)

Using the same grouping and energy range as for the pulsar ($\geqslant 30$ counts per bin, 0.5–7.0 keV), we divided the PWN into several regions (see Figure 2, bottom panel) in order to look for spectral changes (e.g., due to radiative cooling) and to isolate and measure the compact PWN spectrum, which should be less affected by cooling, and hence, offers a better probe of the injected-particle spectral energy distribution (SED).

Region B is an elliptical region (DS9 region format: R.A. = $18^h49^m01\overset{s}{.}6$, decl. = $-00°01'16.''0$, $r_1 = 0.'34$, $r_2 = 0.'19$, $\theta = 304°$) enclosing the brighter inner (more compact) PWN with Region X being excluded. Region X is the circular annulus buffer region mentioned in this section (R.A. = $18^h49^m01\overset{s}{.}7$, decl. = $-00°01'17.''7$, $r_1 = 3.''0$ $r_2 = 0.'13$). No spectrum was extracted from this region. The large-scale extended PWN spectrum is extracted from the elliptical region C (R.A. = $18^h49^m02\overset{s}{.}7$, decl. = $-00°00'24.''9$, $r_1 = 0.'97$, $r_2 = 2.'12$, $\theta = 30°$) with region B being excluded. Several point sources have been also excluded from this extended PWN region. The putative pulsar jet is enclosed by the box D (R.A. = $18^h48^m59\overset{s}{.}9$, decl. = $-00°01'52.''2$, width = $0.'26$, height = $1.'13$, $\theta = 42°$). Two circular background regions ($r = 1.'0$; see Figure 2) were selected nearby, but sufficiently outside of Region D, to minimize the contamination from the extended PWN in a region overlapped by all observations. To quantify the effect of background region selection on the spectral fitting of the large, diffuse emission region enclosed within Region C, different background regions were used. It was found that the choice of background region has a negligible effect on the spectral parameters as differences were within the uncertainties.

Since the PWN emission is relatively faint and significantly contaminated by the background, we fix $n_H$ at the value determined from the fit to the bright pulsar. We note the pulsar spectrum has very few photons below 1.5 keV, and hence, even if it has a thermal soft (see, e.g., Kargaltsev & Pavlov 2007) component unaccounted for in the PL fit, it is unlikely to have any noticeable impact on the determination of $n_H$. The measured large $n_H$ value is consistent with the presence of a dark cloud apparent in optical survey images, e.g., Pan-STARRS (Chambers et al. 2017), which must be located at a 0.5–0.7 kpc distance based on the scarcity of foreground stars and on the 3D dust distribution map (Green et al. 2019). Thus, the cloud is likely in front of the pulsar.

The inner PWN spectrum (shown in Figure 4) is hard with the best-fit $\Gamma = 1.49 \pm 0.20$ and $\chi^2_\nu = 1.09$ for $\nu = 25$. The absorbed and unabsorbed fluxes are $(1.3 \pm 0.1) \times 10^{-13}$ erg cm$^{-2}$ s$^{-1}$ and $(2.7 \pm 0.2) \times 10^{-13}$ erg cm$^{-2}$ s$^{-1}$, respectively. The unabsorbed luminosity is $(1.5 \pm 0.1) \times 10^{33}$ erg s$^{-1}$. The PL fit to the spectrum from a much larger extended PWN region can be described by a PL with $\Gamma = 1.56 \pm 0.11$ with $\chi^2_\nu = 1.0$ for $\nu = 188$. The absorbed and unabsorbed fluxes are $(4.8 \pm 0.2) \times 10^{-13}$ erg cm$^{-2}$ s$^{-1}$ and $11.5^{+0.6}_{-0.8} \times 10^{-13}$ erg cm$^{-2}$ s$^{-1}$, respectively, and the unabsorbed luminosity is $6.7^{+0.4}_{-0.5} \times 10^{33}$ erg s$^{-1}$. Finally, we attempted to fit separately the spectrum from a possible pulsar jet region and obtained a somewhat larger, albeit uncertain, $\Gamma = 1.70 \pm 0.22$. The fit quality is rather poor with $\chi^2_\nu = 2.01$ for $\nu = 16$. Therefore, we cannot say whether the spectrum from the putative jet region is any different from the surrounding emission spectrum.

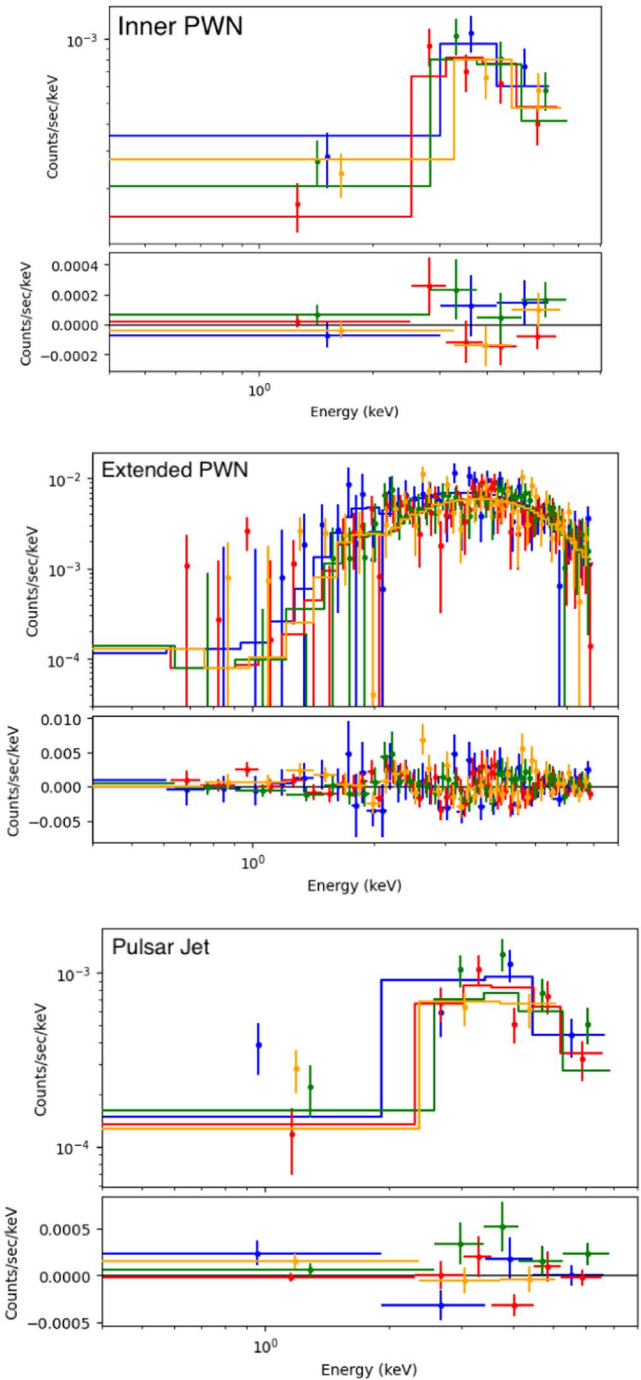

**Figure 4.** Spectra of the PWN around PSR J1849-0001.

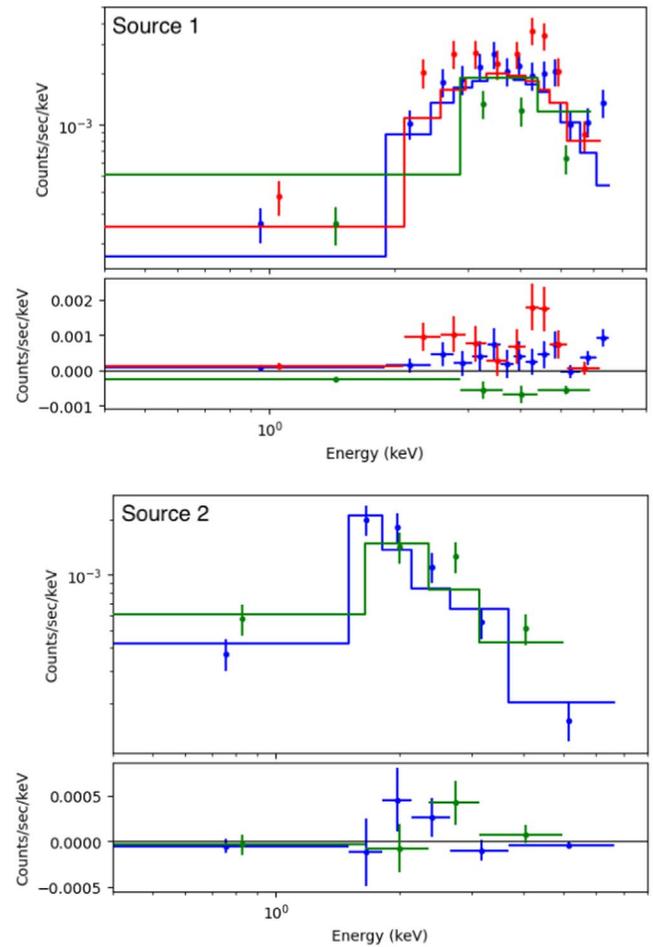

**Figure 5.** Spectra of point sources (Src 1 and 2) found in the CXO ACIS observations (ObsIDs: 13291, 23596, 24494, and 24495). Different colors represent each observation (blue–23596, red–24494, and green–24495).

### 3.4. Brightest Field Sources

We also analyzed the spectra of the two brightest, unresolved X-ray sources shown in Figure 1: Src 1 = CXOU J184829.9-000943 ($\alpha = 282°.1248$, $\delta = -0°.1620$; $1\sigma$ uncertainty $\approx 1.''4$) and Src 2 = CXOU J184851.2+000523 ($\alpha = 282°.2137$, $\delta = 0°.0898$; $1\sigma$ uncertainty $\approx 0.''7$). Both sources were imaged on ACIS-I (ObsIDs 23596, 24494, and 24495) and ACIS-S chips (ObsID 13291). However, due to their variability and different sensitivity of the ACIS-S detector, they are not the brightest in the ACIS-S image. The spectra of both sources (shown in Figure 5) provide a satisfactory fit to an absorbed PL model with $n_H = (3.9 \pm 0.9) \times 10^{22}$ cm$^{-2}$ and $n_H = (8.2 \pm 4.3) \times 10^{21}$ cm$^{-2}$ for Src 1 and Src 2, respectively (see Table 2 for details). Src 1 is strongly variable (see Figure 6). Its flux changed from $(2.9 \pm 0.2) \times 10^{-13}$ erg cm$^{-2}$ s$^{-1}$ to $(1.2 \pm 0.2) \times 10^{-13}$ erg cm$^{-2}$ s$^{-1}$ between 2021 September 26 and 2022 September 11. The source does not have optical/near-infrared counterparts in 2MASS, Gaia DR3, or AllWISE catalogs. However, the more sensitive Pan-STARRS DR1 (Chambers et al. 2017) and UKIDSS DR6 (UKIDSS Consortium 2012) surveys show that Src 1 is located in a fairly crowded environment with several sources within a few

**Table 2**
Spectral Fit Results for PWN Regions and Point Sources

| Region | Name | Area (arcsec$^2$) | Counts | $\Gamma$ | $\mathcal{N}_{-5}$ (10$^{-5}$ photons s$^{-1}$ cm$^{-2}$ keV$^{-1}$ at 1 keV) | $\chi_\nu^2$ ($\nu$ d.o.f.) | $F_{-13}$ (10$^{-13}$ erg cm$^{-2}$ s$^{-1}$) | $F_{-13}^{\text{unabs}}$ (10$^{-13}$ erg cm$^{-2}$ s$^{-1}$) |
|---|---|---|---|---|---|---|---|---|
| a | Pulsar | 28 | 10673 ± 103 | 1.20 ± 0.07 | 46.18 ± 5.04 | 1.03 (311) | $19.72^{+0.24}_{-0.25}$ | $38.70^{+1.33}_{-1.55}$ |
| b | Inner PWN | 531 | 422 ± 22 | 1.49 ± 0.20 | 2.42 ± 0.66 | 1.05 (13) | $0.62^{+0.14}_{-0.12}$ | $1.49^{+0.24}_{-0.29}$ |
| c | Extended PWN | 22476 | 3026 ± 96 | 1.56 ± 0.11 | 19.47 ± 2.93 | 1.0 (188) | 4.78 ± 0.18 | $11.46^{+0.63}_{-0.79}$ |
| d | Pulsar Jet | 1054 | 437 ± 24 | 1.70 ± 0.22 | 3.40 ± 0.99 | 2.01 (16) | $0.66^{+0.05}_{-0.06}$ | $1.78^{+0.20}_{-0.26}$ |
| e | Src 1 (13291) | 706 | 32 ± 7 | ⋯ | ⋯ | ⋯ | 0.26 ± 0.08 | ⋯ |
| e | Src 1 (23596) | 706 | 421 ± 21 | 1.34 ± 0.18 | 7.41 ± 1.83 | 1.37 (12) | $2.62^{+0.18}_{-0.19}$ | $5.37^{+0.37}_{-0.51}$ |
| e | Src 1 (24494) | 706 | 310 ± 18 | 1.60 ± 0.21 | 12.18 ± 3.28 | 1.33 (8) | $2.89^{+0.21}_{-0.24}$ | $6.90^{+0.63}_{-0.83}$ |
| e | Src 1 (24495) | 706 | 126 ± 12 | 1.39 ± 0.36 | 3.78 ± 1.82 | 0.78 (2) | $1.18^{+0.16}_{-0.20}$ | $2.57^{+0.31}_{-0.54}$ |
| f | Src 2 (13291) | 312 | 32 ± 6 | ⋯ | ⋯ | ⋯ | 0.11 ± 0.03 | ⋯ |
| f | Src 2 (23596) | 312 | 168 ± 13 | 2.80 ± 0.19 | 6.71 ± 1.18 | 1.27 (4) | 0.68 ± + 0.06 | 2.06 ± 0.29 |
| f | Src 2 (24494) | 312 | 63 ± 8 | ⋯ | ⋯ | ⋯ | 0.46 ± 0.10 | ⋯ |
| f | Src 2 (24495) | 312 | 121 ± 11 | 2.37 ± 0.25 | 5.35 ± 1.26 | 1.0 (2) | 0.81 ± + 0.09 | $1.90^{+0.28}_{-0.26}$ |

**Note.** PWN spectral fits are performed jointly (all four ACIS observations are included). The different regions of the PWN are shown in Figure 2, bottom panel. Listed are the region name, area, net counts, photon index $\Gamma$, PL normalization $\mathcal{N}_{-5}$, reduced $\chi_\nu^2$ ($\nu$ d.o.f.), and absorbed and unabsorbed 0.5–7 keV fluxes. In all fits for the PWN we set $n_H = 4.40 \times 10^{22}$ cm$^{-2}$. The fits to Src 1 and Src 2 spectra are performed for each ObsID (listed in parentheses) individually because these sources are variable. For Source 1 $n_H = (3.87 \pm 0.90) \times 10^{22}$ cm$^{-2}$, and for Source 2 $n_H = (0.82 \pm 0.43) \times 10^{22}$ cm$^{-2}$. Src 1 (13291), Src 2 (13291), and Src 2 (24494) did not have enough data for a meaningfully constrained spectral fit. The absorbed fluxes of the point sources Src 1 (13291) and Src 12 (13291, 24494) were obtained from the Chandra Source Catalog (v. 2.0.1; Evans et al. 2020).

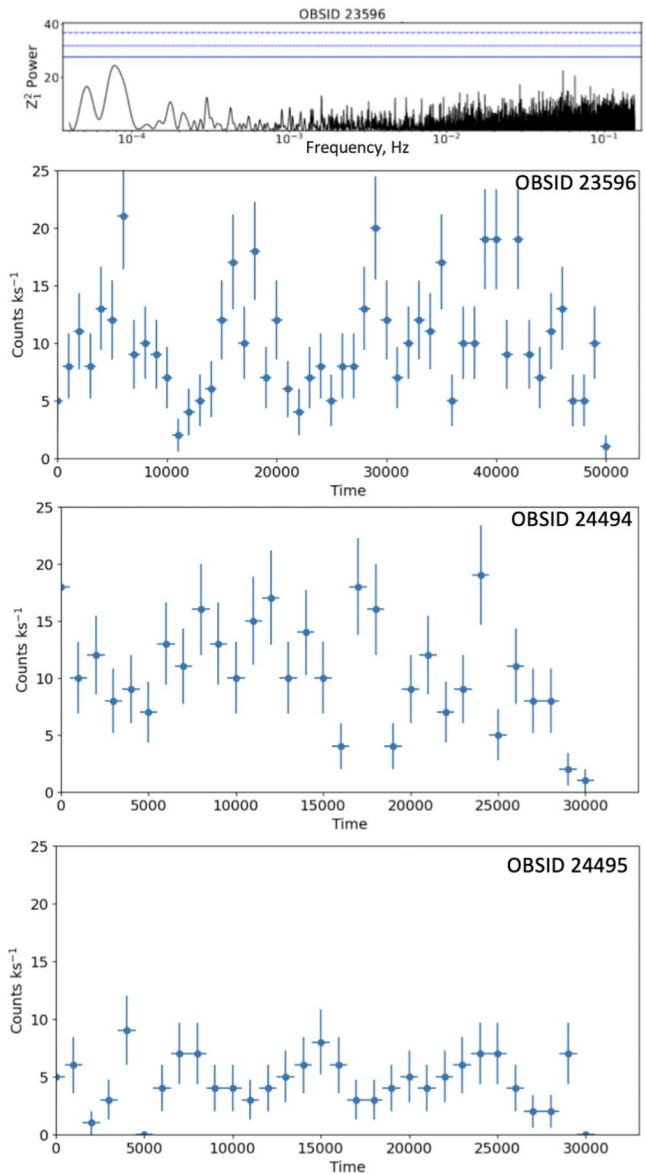

**Figure 6.** Lightcurves of CXOU J184829.9-000943 from three ACIS-I observations (three bottom panels) at 1 ks binning. The top panel shows $Z_n^2$ as a function of frequency. The highest peak, at $\approx 80$ $\mu$Hz, corresponds to the hint of periodicity seen in the first half of the lightcurve from ObsID 23956.

arcseconds' radius. The closest potential counterpart to Src 1 is an optical/IR source detected by Pan-STARRS DR1 (with zmag = 19.4 and ymag = 18.5) and UKIDSS DR6 (with jmag = 16.3, hmag = 15.4, and k1mag = 14.9) separated by 1″.5 from X-ray source. If the association is true, the red color of the counterpart and the X-ray brightness make Src 1 a promising active galactic nucleus (AGN) candidate. Src 1 is not detected in radio surveys including VLASS (Lacy et al. 2020) and RACS (McConnell et al. 2020), but some AGNs are radio quiet/faint. Src 2 is also variable by a factor of 4, with the lowest flux being recorded during ObsID 13291. It varies by a factor of 2 among the three or more recent observations with no clear evidence of a flare within either of the observations. Within its error ellipse, Src 2 is coincident with two Gaia DR3 sources 4266382197008542592 and 4266382196999393920, which are offset by 0″.42 and 0″.55, respectively. Both optical sources have similar proper motion (in terms of direction and magnitude) and are located at the same (within the uncertainties) distance of $\approx$700 pc. Therefore, they may be components of a binary system with an orbital separation of about 420 au. The observed ACIS-I fluxes correspond to the unabsorbed luminosities of $(0.4 - 2) \times 10^{31}$ erg s$^{-1}$. Src 2 is also not detected in radio surveys including VLASS (Lacy et al. 2020) and RACS (McConnell et al. 2020).

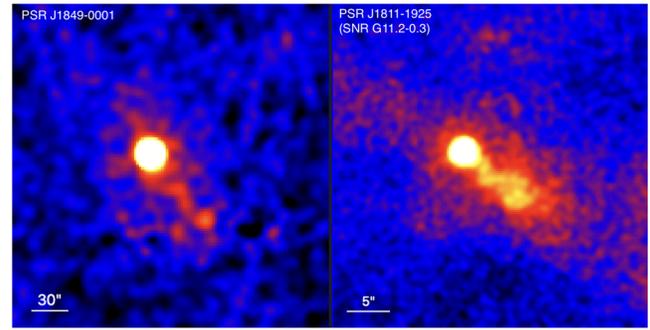

**Figure 7.** Comparison of PWNe produced by the PSRs J1849–0001 and J1811-1925 (in SNR G11.2-0.3). The latter pulsar is also young (∼2 kyr) and energetic ($6.5 \times 10^{36}$ erg s$^{-1}$), with $P = 64$ ms. Its compact PWN appearance is dominated by a helical jet.

For the brighter Src 1, we performed a periodicity search in three ACIS-I observations (individually, after applying barycentric correction with CIAO's axbary) using the $Z_n^2$ test (for the number of harmonics $n = 1, 2$; Buccheri et al. 1983) in the $2T_{\rm obs}^{-1} - 0.15625$ Hz frequency range, where $T_{\rm obs}$ is the observation duration. The frequency resolution was taken to be a factor of 50 smaller than $1/T_{\rm obs}$. No significant periodicity was discovered in any of the observations except for a hint of ∼12 ks periodicity in the data from ObsID 23596 (see the top panel of Figure 6), which is not statistically significant in these data.

## 4. Discussion

The deep ACIS images revealed that the PWN of PSR J1849–0001 has rather amorphous morphology and does not belong to the torus-jet (with the torus being dominant) type of PWNe like those powered by Crab or Vela pulsars. If the faint structure extending SW of the pulsar is indeed a jet, then it may rather resemble PWNe in young Kes 75 or SNR G11.2-0.3 where only one of the two jets is clearly seen (see Figure 7). The one-sidedness can be explained by the pulsar spin axis (also jet direction) pointing close to the line of sight. In this case the jet emission is Doppler boosted, while the counterjet emission is deboosted. The evidence for a jet and the lack of discernible torus also suggest that the inclination angle $\alpha$ between the magnetic and spin axis is small (Bühler & Giomi 2016).

The compact (inner) PWN luminosity (region B in Figure 2) is $\approx 1.6 \times 10^{33}$ erg s$^{-1}$ in 0.5–7 keV (at $d = 7$ kpc), or $\approx 1.6 \times 10^{-4} \dot{E}$, which is below typical X-ray efficiencies of PWNe produced by similarly energetic pulsars (see, e.g., Figure 2 in Kargaltsev et al. 2013). The extended PWN spectrum is softer than that of the inner PWN. The softening can be attributed to the radiative (synchrotron) energy losses, which would then imply a spectral break and a flatter extended PWN spectrum at lower frequencies.

The combined (from regions B and C) X-ray luminosity $L_X \approx 8.3 \times 10^{33}$ erg s$^{-1}$ corresponds to the radiative efficiency $\eta_{X,\rm pwn} = L_X/\dot{E} \approx 8.5 \times 10^{-2}$. In comparison, the HESS

J1849–000 TeV luminosity, $L_{\rm TeV} \approx 1.2 \times 10^{34}$ erg s$^{-1}$ (in 1–10 TeV), corresponds to the TeV radiative efficiency of $1.2 \times 10^{-3}$, which are fairly typical values for TeV PWNe (H.E.S.S. Collaboration et al. 2018a). We note that, unlike X-ray efficiency, the TeV efficiency is not contemporaneous because the cooling time of TeV-emitting particles is about a factor of 10 longer (at 1 TeV), and hence, the pulsar's $\dot{E}$ may have been significantly larger when the TeV-emitting particles were injected. Amenomori et al. (2023) measured the spectrum of HESS J1849-000 at TeV energies and found no evidence of cutoff up to $E_\gamma = 740$ TeV. For the inverse-Compton (IC) on cosmic microwave background (CMB) photons,[10] this implies the corresponding electron energy of $E_e \approx 2.15 (E_\gamma/1\,{\rm PeV})^{0.77} = 1.7$ PeV (Amenomori et al. 2023). This energy corresponds to ∼30% of the energy, $e(\dot{E}/c)^{1/2}$, corresponding to the current potential drop across the polar cap of PSR J1849-0001.

Since HESS J1849–000 is relatively compact compared to some other TeV sources associated with PWNe, we attempted to model the multiwavelength spectrum with a simple one-zone (leptonic) radiation model using the Python package naima (Zabalza 2015). We first modeled the electron SED with an exponential cutoff PL $N(E) = A(E/E_0)^{-p} \exp(-(E/E_{\rm max})^\beta)$, where $A$ is the normalization (in units of particles GeV$^{-1}$ at 1 GeV), $E_0$ is the reference energy (1 GeV), $p$ is the slope of the electron SED, $E_{\rm max}$ is the maximum particle energy, and $\beta$ is the cutoff exponent (which we set to 10 to quickly kill off the SED above the maximum energy). The slope of the electron SED, $p = 2.04$, is inferred from the photon index, $\Gamma = 1.52$, for the absorbed PL model fitted to all PWN regions combined together assuming optically thin synchrotron emission.

The observed X-ray spectrum requires a maximum electron energy of at least $E_{\rm max} = 600$ TeV. The minimum electron energy, chosen to be $E_{\rm min} = E_0 = 1$ GeV (the Naima default) is not known. However, it does not correlate with other model parameters and does not change the fit as long as it is small enough. For the IC spectrum, we use the background photon fields from Popescu et al. (2017): the CMB, a far-IR component ($T = 30$ K, $U = 2 \times 10^{-12}$ erg cm$^{-3}$), a near-IR component ($T = 500$ K, $U = 4 \times 10^{-13}$ erg cm$^{-3}$), and a visible component ($T = 5000$ K, $U = 1.5 \times 10^{-12}$ erg cm$^{-3}$). We varied the values of $B$, $E_{\rm max}$, and normalization $A$ until we obtained a reasonably good match between the model and the data (see Figure 8). However, the inferred value of the magnetic field, $1-2$ $\mu$G, is low even when compared to the typical Galactic plane field (∼5 $\mu$G; Ferrière 2015). Decreasing the TeV fluxes by a factor of 2–3 (an estimate of the TeV flux for the X-ray-emitting PWN volume) only allows for a ∼50% increase in the magnetic field strength. This may be an artifact of using a simplistic one-zone model that does not include the expected spectral evolution, causing the SED of older particles to be different from that of the freshly injected ones (e.g., due to the radiative cooling), nor does it account for a possible large variation of $B$ with the distance from the pulsar or a variable particle injection rate of particles throughout the life of the pulsar. We note that the radiative losses are expected to produce a cooling break in the spectrum (see, e.g., discussion of cooling breaks in Klingler et al. 2018). Although the location of the break is currently unconstrained (the radio upper limit is too high; see Figure 8), placing it at a reasonable energy (based on other young PWN, e.g., the Crab, Vela, or Mouse PWN spectra where such a break is constrained) does not affect the inferred values of the magnetic field and $E_{\rm max}$. We also note that, for this source, the simplistic one-zone model could perform better here than for some other large TeV/X-ray PWNe (e.g., Li et al. 2023) because for HESS J1849-000, the TeV source size is comparable to the X-ray PWN size (see Figure 9). Despite this, the simplistic one-zone model may be lacking in that it is not accounting for the presence of an older "relic" electron population that is no longer emitting synchrotron emission in X-rays but which could still be emitting inverse-Compton emission in TeV $\gamma$-rays (as IC cooling timescales, ∼10 kyr, are an order of magnitude longer than synchrotron cooling timescales, ∼1 kyr, for typical PWN parameters). Finally, we note that although Amenomori et al. (2023) discussed an alternative hadronic scenario, where the TeV $\gamma$-rays are produced by neutral pion decay, no evidence of a radio SNR is seen in the radio images (see Figure 9). The paper of Kim et al. (2024), which appeared nearly simultaneously with our paper, offers a more detailed modeling of multiwavelength emission from this source, including the comparison between one-zone and more physical multizone models.

---

[10] Upscattering of CMB photons makes the largest contribution (compared to other background photons) at these very high energies (Amenomori et al. 2023).

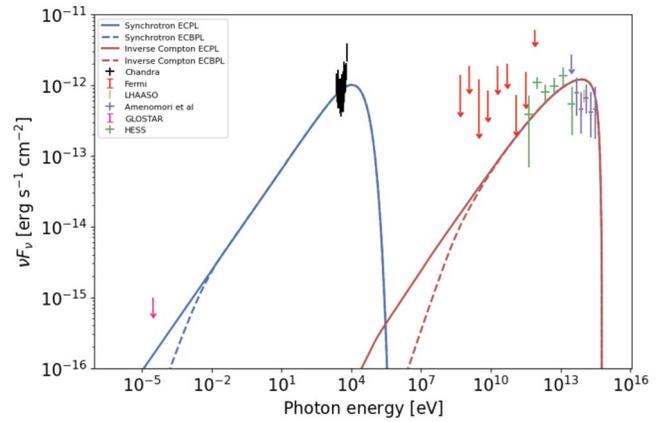

**Figure 8.** Multiwavelength SED of the J1849 PWN. Shown are the data from Chandra (this paper), Fermi-LAT (upper limits; Eagle 2022), HESS (H. E. S. S. Collaboration et al. 2018b), LHAASO (Cao et al. 2021), and the Tibet air shower array (TASA; Amenomori et al. 2023). The curves show synchrotron (blue) and IC (red) emission from an exponential cutoff PL (ECPL; solid) and an exponential cutoff broken PL (ECBPL; dashed) electron SED in a one-zone PWN model. The $2\sigma$ GLOSTAR survey limit (at 6.9 GHz; Brunthaler et al. 2021) is shown, assuming the radio source to be as large as the TeV source. The survey's angular resolution is 18″. Note that the survey's largest resolvable angular scale is 4′, which is a factor of 2 smaller than the TeV source size (see Figure 9). Therefore, the upper limit is very approximate. The limit from the Very Large Array (VLA) Global Positioning System (GPS) is significantly higher (not shown for this reason) although GPS can resolve larger angular scales than the GLOSTAR survey.

For the pulsar, $L_X \approx 2.3 \times 10^{34}$ erg s$^{-1}$, and the radiative efficiency $\eta_{X,\rm psr} \approx 2.3 \times 10^{-3}$. The latter is in line with values typically seen in young pulsars. It is important to note that PSR J1849–0001 has not been detected in radio or in $\gamma$-rays. Among 65 pulsars with $\dot{E} > 10^{36}$ erg s$^{-1}$, only eight pulsars are both radio and $\gamma$-ray quiet. In fact, there is only one pulsar with $\dot{E}$ higher than that of PSR J1849–0001 that is radio and $\gamma$-ray quiet (J1813–1749 in SNR 12.8–0.0; also an INTEGRAL source). It is a bit of a mystery why some young, high-$\dot{E}$ pulsars exhibit pulsations in X-rays/soft $\gamma$-rays but are quiet (or very weak) in radio and $\gamma$-rays. It is not that particle acceleration and pair creation are inefficient in these pulsars because their X-ray luminosities are not unusual. A more likely explanation is a specific magnetospheric geometry—the combination of angle ($\alpha$) between the magnetic dipole and spin axis and/or an unfavorable viewing angle ($\zeta$)—makes the narrow radio beam and the $\gamma$-ray emission from the outer magnetosphere miss the Earth, while the broader beam of nonthermal X-ray emission from pairs above the polar cap is still seen. In this respect, a similar case could be that of another $\gamma$-ray- and radio-quiet pulsar PSR J1811-1925 (Torii et al. 1997) where deep CXO observations (Borkowski et al. 2016) also revealed an underluminous compact PWN with clear evidence of jet but only a weak indication of a torus. The 3–8 keV pulse profiles of PSR J1849–0001 (Bogdanov et al. 2019) and PSR J1811-1925 (Madsen et al. 2020) are also similar, showing a single broad peak per rotation period. One of the possible consequences of such a geometry is the relative faintness of the PWN because, for a more aligned rotator, a smaller volume is occupied by the striped pulsar wind zone where the magnetic field energy is converted to the energy of pulsar wind particles. We also note that pulsars having small $\alpha$ are expected to be weak $\gamma$-ray sources lacking pulsations (see, e.g., Cerutti et al. 2016).

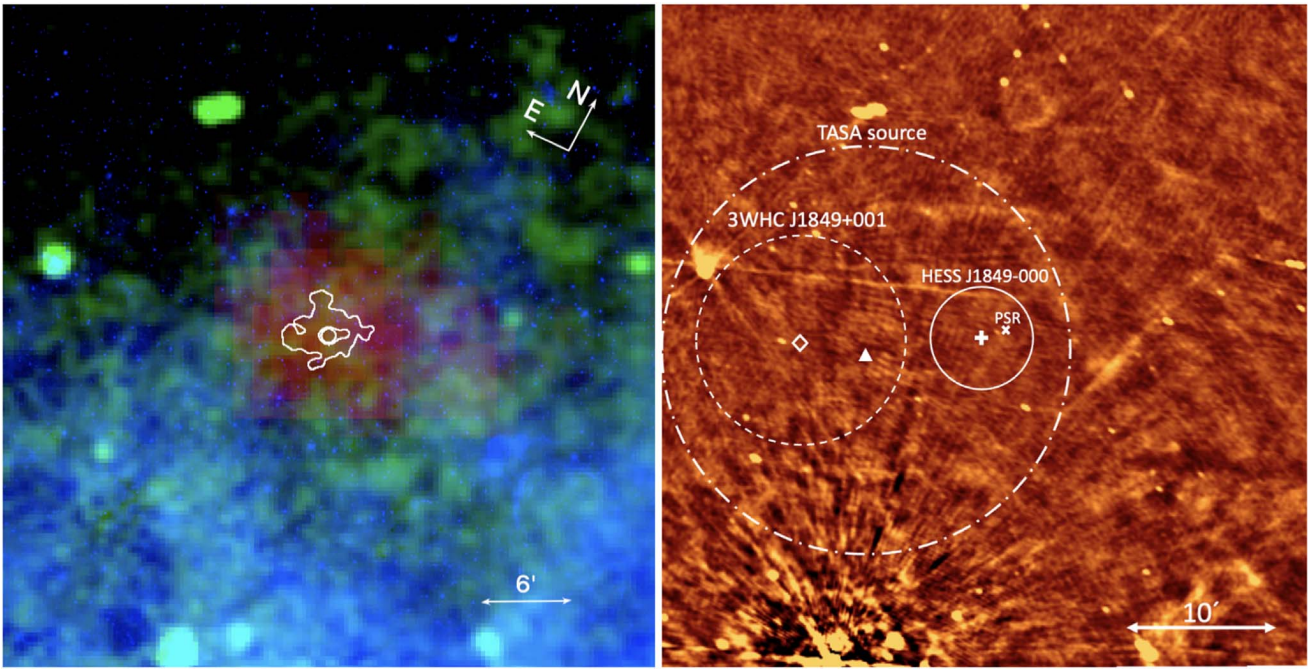

**Figure 9.** The left panel shows the true-color image: red–TeV from HESS GPS (H. E. S. S. Collaboration et al. 2018b), green–1.42 GHz continuum radio from VLA Galactic Plane Survey (Stil et al. 2006), blue–Spitzer IRAC 8 micron image from GLIMPSE 360 survey (Churchwell 2009). The contours are X-rays based on the image shown in the top panel of Figure 1. The right panel is the 6 GHz radio image from the GLOSTAR survey (Brunthaler et al. 2021), showing the same area of the sky as in the left panel. For the HESS source, the solid circle shows the extent of the source (H. E. S. S. Collaboration et al. 2018b). For the 3HWC source, the dashed circle shows the positional uncertainty (systematic and 1σ statistical combined; Albert et al. 2020). The dashed–dotted circle shows the positional uncertainty of the very high-energy source detected by the TASA (systematic and 1σ statistical combined; Amenomori et al. 2023).

## 5. Conclusions

CXO observations show that young and energetic PSR J1849–0001 is surrounded by an underluminous compact PWN, which lacks the often-seen torus-jet morphology. The combined CXO image shows amorphous PWN morphology with some evidence of a jet and no evidence of a torus. Morphologically, the PWN of PSR J1849–0001 may resemble the brighter PWN of PSR J1811–1925. Both of these pulsars feature hard nonthermal spectra, are $\gamma$-ray and radio quiet, have similar pulse profiles in X-rays, and also have similar spin-down properties. Both pulsars also belong to the so-called MeV pulsar class, which could have small inclination angles between the spin and magnetic dipole axis.

The fainter extended X-ray PWN is discernible up to a few arcminutes in the combined CXO images. This PWN is one of the few where the connection between the TeV (HESS J1849–000) and X-ray PWNe is credible because of the good positional coincidence and relative compactness of the TeV source. The comparable sizes of the X-ray and TeV PWNe could make the simplistic one-zone modeling more relevant than in some other X-ray/TeV PWN cases. However, the inferred magnetic field value is still suspiciously low, suggesting that even in such cases, the dynamics of particle injection and varying magnetic fields are likely to have a large impact on the observed PWN emission.

We also find a curiously bright and variable X-ray source, which is apparently associated with a very wide binary that is not expected to show any additional X-ray activity due to the large separation between the stars. The X-ray lightcurves do not show flares typical of coronally active stars. Optical spectroscopy could help to better understand the type of star(s) and underlying reasons for X-ray activity.

## Acknowledgments

The authors are grateful to Andrew Sargent for the discussion of the radio survey images. Support for this work was provided by the National Aeronautics and Space Administration through the National Aeronautics and Space Administration through Chandra Award No. GO1-22073X issued by the Chandra X-ray Center, which is operated by the Smithsonian Astrophysical Observatory for and on behalf of the National Aeronautics Space Administration under contract NAS8-03060. J.H. and N.K. acknowledge support from NASA under award No. 80GSFC21M0002.

This paper employs a list of Chandra data sets, obtained by the Chandra X-ray Observatory, contained in the Chandra Data Collection (CDC) 215 doi:10.25574/cdc.215.

*Software:* CIAO v4.15 (Fruscione et al. 2006), Naima (Zabalza 2015), Sherpa (Freeman et al. 2001), Wavdetect (Freeman et al. 2002)

## Appendix
## Astrometry

We perform both relative astrometry by aligning X-ray sources from individual observations and absolute astrometry by tying X-ray source positions in the merged image to optical counterparts in the Gaia DR3 catalog (Gaia Collaboration et al. 2023).

The CIAO tool `wcs_match` is used to determine the translational shift minimizing the sum of positional offsets between the matched pairs. The $1\sigma$ uncertainty of the astrometric correction is calculated using the WRMSR as

$$\text{WRMSR}^2 = \frac{\sum_{i=1}^{N} w_i R_i^2}{\sum_{i=1}^{N} w_i} \tag{A1}$$

with

$$w_i = \frac{1}{\sigma_i^2} = \frac{1}{\sigma_{\text{ref},i}^2 + \sigma_{\text{X},i}^2}, \tag{A2}$$

where $R_i$ is the residual for the $i$th pair after the astrometric correction and $\sigma_{\text{ref},i}$ and $\sigma_{\text{X},i}$ are the positional uncertainties of the reference source and the X-ray source to be aligned with the $i$th pair.

To find optimal parameters for `wcs_match`[12] and ensure the reliability of matched pairs for astrometric correction, we perform `wcs_match` on a grid of search radii and residual limits. We select values yielding ⩾5 matched pairs, with astrometric correction (and its uncertainty WRMSR) consistent with most solutions from the parameter space grid search. For relative astrometry, we use significant X-ray sources (detection significance $>3\sigma$ and net counts $>10$) detected from `wavdetect` (Freeman et al. 2002), excluding the pulsar, to coalign the reference X-ray image (ObsID=23596) and the other three images. Corrected (with `wcs_update`) images are coadded to create a deeper merged image using CIAO task `merge_obs`. For absolute astrometry, a more stringent selection ($\sigma > 5$ and net counts $>20$) is applied to X-ray sources detected from the merged image before aligning them to Gaia DR3 sources. Finally, we update the WCS information and the aspect history files accordingly using `wcs_update`. The relative and absolute astrometry solutions are summarized in the first and second parts of Table 3.

## ORCID iDs

Seth Gagnon https://orcid.org/0000-0003-0902-1935
Oleg Kargaltsev https://orcid.org/0000-0002-6447-4251
Noel Klingler https://orcid.org/0000-0002-7465-0941
Jeremy Hare https://orcid.org/0000-0002-8548-482X
Hui Yang https://orcid.org/0000-0002-8832-6077
Jordan Eagle https://orcid.org/0000-0001-9633-3165

**Table 3**
The Relative Astrometry Solutions in the First Part of the Table with the Reference ObsID = 23596 and the Absolute Astrometry Solution (Corrected to Gaia DR3) in the Second Part of the Table

| ObsID | Radius[a] | Reslim[b] | ΔR.A. (arcsec) | Δdecl. (arcsec) | WRMSR (arcsec) | No. of pairs |
|---|---|---|---|---|---|---|
| 13291 | 0.6 | 0.45 | −0.23 | 0.40 | 0.18 | 7 |
| 24494 | 1.6 | 0.5 | −0.21 | 1.36 | 0.27 | 6 |
| 24495 | 0.7 | 0.6 | 0.01 | −0.16 | 0.28 | 9 |
| Merged | 1.0 | 0.4 | 0.07 | −0.25 | 0.27 | 6 |

**Notes.**
[a] Source match radius for wcs_match.
[b] Residual limit for wcs_match.

---

[12] See https://cxc.cfa.harvard.edu/ciao/ahelp/wcs_match.html for details.